# Alternative Fertilizer Production Using an Environmentally Benign Approach

Samuel Alpert[1], Visal Veng[2], and Benard Tabu[2],
[1] Department of Electrical and Electronic Engineering, University of Massachusetts Lowell, United States of America
[2] Department of Mechanical Engineering, University of Massachusetts Lowell, United States of America

**Abstract**

Fertilizer manufacturing is an energy-intensive process that is done in a select number of facilities across the world. We performed a literature review of methods that are being pursued in the emerging field of sustainable fertilizer manufacturing and analyzed the results of one such sustainably sourced fertilizer process. The results of our testing on a rooftop garden showed that the production of tomatoes was similar to the control and commercial fertilizer scenarios in that the total weight of tomatoes per plant was 16% higher, and the weight per tomato was 68% more than the control (tap water). We hope that this type of research will continue to be performed and expanded so that we can adopt a more sustainable method of fertilizer manufacturing in our pursuit of net positive environmental benefits.

## 1) Introduction

Fertilizer is required to sustain global food production to meet the constantly growing population, which is projected to be 9.7 billion in 2050[1]. To meet the increasing demand, synthetic fertilizers, including nitrogen (N), phosphorous (P), and potassium (K), are prevalent in the commercial farming industry to ensure an adequate food supply. Nitrogen fertilizer is the main nutrient for plant growth which accounts for over 50% of global synthetic fertilizers globally [2]. However, when nitrogen fertilizer is applied to plants, only 30-40% of nitrogen is absorbed by the plants [3], and the rest is wasted or drained into the water, causing water pollution, so-called "nitrogen runoff" [4]. Currently, the conventional Haber-Bosch process [5] requires high temperature (400° to 650° C) and high pressure (200 to 400 atm) to produce fertilizer, releasing a vast amount of $CO_2$ emissions, approximately 1.4% of global carbon dioxide emissions. Being a very energy-intensive process, it consumes almost 2.0 % of global energy production.

To alleviate these issues, the use of solar energy and plasma technology to produce fertilizer from water and air is one alternative solution that is gathering attention from researchers around the world. Plasma is an ionized gas consisting of electrons, ions, neutrals, and molecules [6]. There are two kinds of plasma: thermal and nonthermal plasma. In thermal plasma, both the electrons and heavy species have uniform high temperatures, usually from 6000 to 20,000 K [7-11]. However, nonthermal plasma exhibited higher electron temperature (~11,000 or higher) compared to the background gases. The growing interest in non-thermal plasma is attributed to its higher energy efficiency and selectivity. Many studies on the production of nitrogen fertilizer using nonthermal plasma from water and air have been conducted [12, 13]. However, the use of solar as an energy source to generate the transfer-arc plasma mode in this process as well as the in-situ production of fertilizer in a distributed manner, is a novel approach that may help smaller farms, communities, and countries that do not have access to large fertilizer manufacturing plants utilizing the Haber-Bosch process [14]. Considering the decreasing cost and technological advancement in renewable energy, particularly solar and wind, it has gained interest not only in the US but worldwide [15]. With the proposed approach, nitrogen fertilizer is produced from air and water using solar and plasma technology that can supply directly to the plants onsite and on demand.

The research aims to fulfill the increasing need for nitrogen fertilizer at a low-cost, decentralized manner while mitigating nitrogen pollution and carbon dioxide ($CO_2$) emissions commonly associated with fertilizer production. Small and medium farmers can benefit from adapting their existing drip irrigation systems to this modular system. The technology tackles three main issues: first, this technology utilizes solar energy as a renewable energy source helping mitigate climate change and pollution from the Haber-Bosch fertilizer production process [16]. Second, it minimizes nitrogen runoff where chemical fertilizer is over-utilized, causing water pollution and is harmful to the ecological system [4]. Finally, it is economically viable in a way that lowers expenses and increases food production, contributing significantly to global food security and sustainable development. This system produces fertilizer on-site, eliminating transportation costs [16].

## 2) Research Methodology

Solar-generated plasma nitrogen fertilizer system is built and tested on roof-top garden. The system produces nitrogen-based fertilizer from air and water using low-temperature atmospheric pressure plasma generated by solar electricity. A sample image of plasma used to produce fertilizer is shown in Figure 1a.

The PV module generates electricity which is then converted to an AC via an inverter. The AC electrical voltage is scaled to over 20 kV by the power supplies. The high voltage then creates the air plasma, which impinges on the water generating nitrates, nitrites, and other species in water. The pilot demonstration was carried out at the rooftop garden of the South Campus of the University of Massachusetts Lowell between May and September of 2022. Three groups of tomatoes were fed with green fertilizer, Fish (commercial fertilizer), and water immediately after transplanting them. The growth in terms of height was monitored. Green Fertilizer and Fish groups were treated with 20 ppm/20ml of green fertilizer and fish respectively weekly for eight weeks, and the third group was treated with tap water of the same amount of 20ml. The groups treated with both fish and water served as control groups.

## 3) Results of our pilot test

We installed a pilot system in an urban garden setting with individual tomato plants, shown in Figure 1b, in the spring of 2022. We have tested out the output and viability of our fertilizer production system for the past 6 months or so. The results show that our system utilizes a minimal amount of stored solar and is easily able to support the automated operation of the system to create fertilizer to be ready on demand. The plant growth was comparable to the commercially available fertilizer applied to a control group. Vegetable production for the test fertilizer was higher than in the comparison and control groups. Nitrates ($NO_3$) content was compared and had positive results. Soil nitrates did not increase after our test fertilizer application. Overall, these results were positive and indicated the possibility to both increase our concentration of fertilizer applied as well as to scale up the number of plants being tested to see if we maintain similar results when scaling up the system.

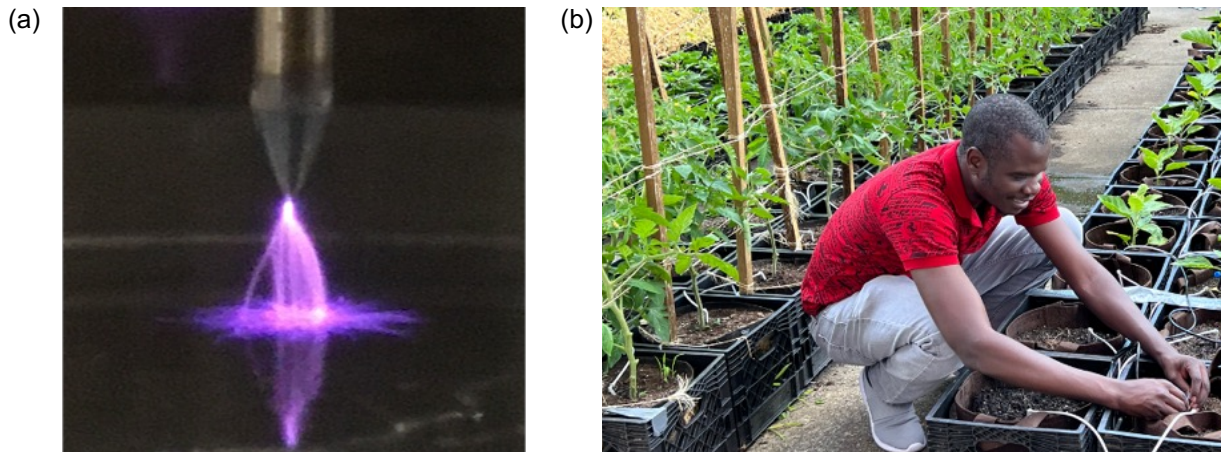

Figure 1: **Plasma production of nitrate fertilizer** (a) plasma close-up view (b) on-site installation and testing

### 3.1 Plant Growth and Vegetable Production

Plant growth was measured, starting from seedling-size tomatoes, and the heights were measured. However, due to issues with the stalks leaning over, we were unable to achieve accurate height comparison results.

When the tomatoes turned a bright red color, we harvested the tomatoes from the three comparison groups, control (tap water), fish fertilizer, and sustainable fertilizer (our own test fertilizer solution). The measurements of weights were performed over a period of about 4 weeks of harvests and resulted in positive results where the test fertilizer solution provided more weight per tomato plant on a normalized basis. See Fig. 3 for the totals which resulted from our tomato weight measurements over the lifecycle of the tomato plants:

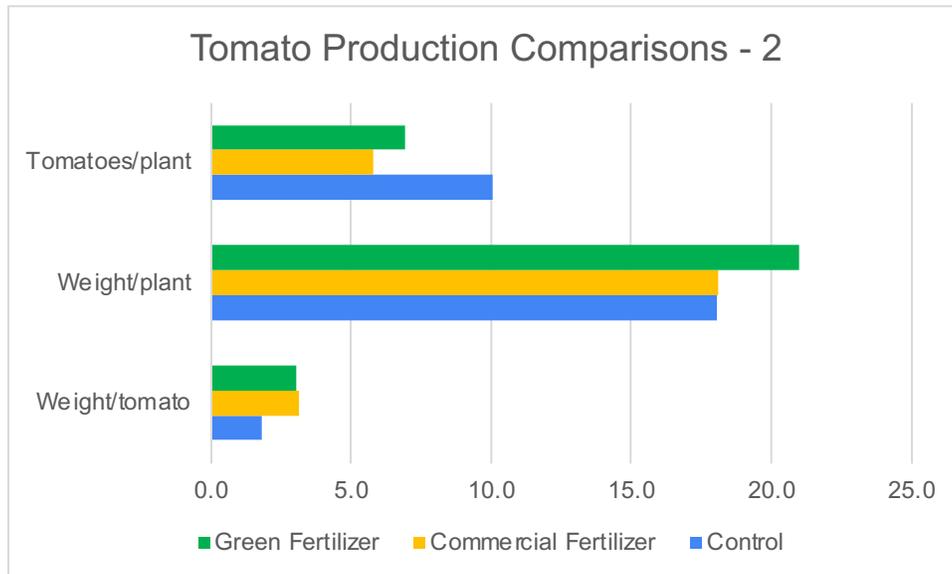

Figure 2: Tomato Production Comparisons

## 3.2 Nitrates Content

We measured the nitrates ($NO_3$) content of commercially bought tomatoes, both medium tomatoes and cherry tomatoes, and compared these results to tomatoes grown with commercial fertilizer, tap water (control), and sustainable fertilizer. The results are shown in the table below, and they appear to indicate that the nitrates content of tomatoes grown with sustainably manufactured fertilizer is comparable to or lower than the nitrates found in the other tomatoes tested.

Table 1: Nitrate Measurements in the Tomatoes Harvested

|  | Nitrates Measured | Compared to Control |
|---|---|---|
| Garden Tomatoes (Tap Water) - CONTROL | 57 | N/A |
| Garden Tomatoes (Fish Fertilizer) | 67 | 118% |
| Garden Tomatoes (Sustainable Fertilizer) | 63 | 111% |

## 3.3 Soil Impacts

We tested soil samples prior to fertilizer application and mid-season, and the samples showed a decrease in nutrients after the growing season had begun. This could imply that there are some nitrates leaching from both the fertilized soils as well as the tap water cases. However, the amount of nitrates from the control case being more than the green fertilizer scenario could mean that the nitrates in tap water are less able to be absorbed directly by the plant after the season has begun than the tomatoes that were fed with our test "green fertilizer."

Table 2: Nitrate and pH Results of the Soil

|  | Nitrates Measured Pre-season | Nitrates Measured Mid-season | pH Pre-season | pH Mid-season |
|---|---|---|---|---|
| Garden Tomatoes (Tap Water) - CONTROL | 23 | 75 | 7.1 | 5.9 |
| Garden Tomatoes (Fish Fertilizer) | 23 | 54 | 7.1 | 6.0 |
| Garden Tomatoes (Green Fertilizer) | 23 | 32 | 7.1 | 6.2 |

### 3.4 Social and Environmental Impacts

The research can provide significant social benefits, particularly relevant to key stakeholders involved in the agricultural sector ranging from economic, health, local community improvement to promoting clean and green farming practices. For farmers, the system is expected to generate more income for them in the long run because fertilizer is produced on-site and on demand, so there is no transportation cost associated with this approach, and there is no labor cost when it comes to fertilizer supply. Moreover, this system is designed to have a 15-year lifespan based on the lifespan of solar panels and other system components. The system will help remote or indigenous farmers without access to an electrical grid or fertilizer supply. With access to low-cost fertilizer production systems, their crop yield can be increased, promoting local communities economically. The resiliency of farmers in the community allows them to give back to their community. The system design provides clean and green fertilizer for farmers, so their health could also be enhanced, especially in developing countries where in most cases, farmers rely on manure or compost to supply to their agricultural farm, which is labor and time-consuming and affects their health through the process. The system produces organic fertilizer because it is a natural chemical process from the sunlight to liquid fertilizer, and no chemical substances are being added to the system, promoting clean and green farming practices. A system can enhance the livelihoods of communities most in need of support.

### 3.5 Environmental Impacts

This approach is a carbon-free process, positively contributing to clean air. The current artificial fertilizer production method accounts for over 300 MMT of annual $CO_2$ emissions in the agriculture section alone. The system can mitigate nitrogen runoff (water pollution) resulting from the over-utilization of fertilizer when plants cannot absorb all fertilizer and go into water sources through the rain. The system promotes using renewable energy, such as solar and wind energy, which is clean and environmentally benign. The environmental benefits of utilizing solar electricity instead of natural gas to generate nitrogen fertilizer are dramatic and result in a 95% reduction of greenhouse gas emissions compared to traditional methods based on sourcing recycled solar panels. Recycled solar panels, since they do not require much additional energy usage except for repairing and testing the meetings, the greenhouse gas emissions are dramatically reduced to almost zero compared to manufacturing solar panels from entirely raw materials.

## 4) Conclusion

To dramatically reduce our environmental emissions in the agricultural sector and become closer to carbon neutral or net zero emissions, we must first shift how we access raw materials such as fertilizer. Fertilizer manufacturing needs are growing worldwide, and to become stewards to an environmentally friendly method of agricultural and fertilizer needs, there are options to move toward, and changes to make that will enhance farmers' resiliency and reduce the environmental impact. Research in fertilizer manufacturing at a distributed level will help make this happen. As the preliminary results of our experiments show, using a sustainably produced fertilizer is on par with using traditional fertilizer, even without added phosphorous and potassium. This sustainable fertilizer pathway will lead us to a better more for ourselves and the next generation due to reduced environmental emissions and reduced fluctuation of fertilizer costs.

**Acknowledgments**

We want to thank the Sustainability Office of the University of Massachusetts Lowell for allowing us to install our pilot research installation and assisting with funding for our research. Our appreciation goes out to the Rist DifferenceMaker Institute for our first grant award that kicked off our research into the commercialization world. We sincerely thank our advisors, Professor Juan Pablo Trelles and Professor Cordula Schmid, for guiding our research efforts. We also would like to acknowledge the assistance of Mill City Grows and especially Brian Mariano, who was instrumental in aligning us with farming principles and helping us foresee the benefits of our research to support the urban farming community.